\documentstyle[preprint,aps,tighten,epsf]{revtex}

\begin{document}
\preprint{\vbox{\baselineskip=12pt
\rightline{SNUTP 97-045}
\rightline{APCTP 97-007} }}

\draft
\title{
The Schwarzschild Solution in the 4-Dimensional \\
Kaluza-Klein Description of The Einstein's Equations}
\author{ J.H. Yoon\cite{email}, 
S.K. Oh, C.M. Kim}
\address{
Department of Physics and Institute for Advanced Physics\\
Kon-Kuk University, Seoul 143-701, Korea}
\author{ Y.M. Cho}
\address{ Asia Pacific Center for Theoretical Physics\\
and\\
Department of Physics, Seoul National University, Seoul 151-742, 
Korea}

\maketitle
\thispagestyle{empty}

\begin{abstract}

The Kaluza-Klein formalism of the Einstein's theory, based on  
the (2,2)-fibration of a generic 4-dimensional spacetime, describes
general relativity as a Yang-Mills gauge theory on the 
2-dimensional base manifold, where the local gauge symmetry
is the group of the diffeomorphisms of the 2-dimensional fibre 
manifold. 
As a way of illustrating how to use this formalism in finding exact 
solutions, we apply this formalism to the spherically symmetric case, 
and obtain the Schwarzschild solution by solving the field equations. 
\end{abstract}


In spite of those efforts that have been made over the decades 
trying to understand general relativity as a local gauge theory 
such as the Maxwell or a Yang-Mills theory, it seems 
fair to say that the proper gauge theory formulation 
of general relativity is still lacking. 
Such a gauge theory formulation, if feasible at all, would allow us
to understand general relativity in terms of familiar notions
of standard gauge theories. 
Recently, we have proposed a Kaluza-Klein 
formalism\cite{propose,four} of general relativity, 
based on the (2,2)-fibration\cite{hayward,d'inverno} of a generic 
4-dimensional spacetime. In this (2,2)-fibration, 
the 4-dimensional spacetime is regarded as a local product
of a 2-dimensional base manifold and a 2-dimensional fibre 
manifold. Introducing 
the Kaluza-Klein variables adapted to this fibration, we found that
general relativity of 4-dimensional spacetimes can be written
as a Yang-Mills gauge theory defined on the 2-dimensional 
base manifold, where the local gauge symmetry is
the group of the diffeomorphisms of the 2-dimensional fibre
manifold. The appearance of the diffeomorphisms of the 2-dimensional 
fibre space as the Yang-Mills gauge
symmetry, among others, is the most distinguished feature 
of this formalism, which is valid for a generic spacetime 
that does {\it not} possess any isometries whatsoever. 

In this Letter, as an application of this formalism, 
we shall obtain the Schwarzschild solution by solving the Einstein's 
equations written in the Kaluza-Klein variables. 
After a short introduction to the general formalism\cite{summary}, 
we shall present the Einstein's equations written in the Kaluza-Klein 
variables. Then we shall solve the field equations, 
using the spherical symmetry. Discussions on possible applications 
of this formalism will follow. 

Let us start by recalling that 
any metric of a 4-dimensional spacetime of 
the Lorentzian signature may be put 
to the following form\cite{summary}
\begin{equation}
ds^2 
= -2dudv - 2hdu^2 +{\rm e}^{\sigma} \rho_{ab}
    \left( dy^a + A_{+}^{\ a}du +A_{-}^{\ a} dv \right)
   \left( dy^b + A_{+}^{\ b}du +A_{-}^{\ b} dv \right), \label{pol}
\end{equation}
at least locally, 
where $\rho_{ab}(a,b=2,3)$ is the conformal 2-geometry 
of the transverse surface 
on the hypersurface $u={\rm constant}$, satisfying the condition 
\begin{equation}
{\rm det}\ \rho_{ab}=1.                        \label{det}
\end{equation}
The geometry represented by the above metric 
can be best understood in terms of the {\it bundle} geometry;
$(u,v)$ are the coordinates of the 2-dimensional base manifold 
denoted by $M_{1+1}$, and $y^{a}$ are 
the coordinates of the spacelike 2-dimensional fibre space denoted by
$N_{2}$. From (\ref{pol}) we find that 
the covariant metric of $M_{1+1}$ is given by
\begin{eqnarray}
\left( \begin{array}{rr}
                       -2h & -1 \\
                       -1  &  0 
       \end{array}\right).      \label{poly}
\end{eqnarray}
The field $\sigma$ is a measure of the area of $N_{2}$,
and the fields $A_{\pm}^{\ a}$ are the connecting vector fields
that define the horizontal lift vector fields 
orthogonal to the fibre space. Notice that
$h$, $\sigma$, $\rho_{ab}$, and $A_{\pm}^{\ a}$ in (\ref{pol})
are functions of all the coordinates ($u,v,y^{a}$), as we assume no
spacetime isometries. 

The equations of motion of $h$, $\sigma$, $\rho_{ab}$, 
and $A_{\pm}^{\ a}$ can be obtained by varying $I_{4}$,
the 4-volume integral of the scalar curvature $R_{4}$ 
of the spacetime represented by the metric (\ref{pol}), 
which is given by
\begin{eqnarray}
I_{\rm 4}& = &  \int \! \! du \, dv \, d^{2}y \, 
   {\rm e}^{\sigma}R_{4}           \nonumber\\
&=& \int \! \! du \, dv \, d^{2}y \,
\Big[ -{1\over 2}{\rm e}^{2 \sigma}\rho_{a b}
  F_{+-}^{\ \ a}F_{+-}^{\ \ b}
  +{\rm e}^{\sigma} (D_{+}\sigma) (D_{-}\sigma) 
  -{1\over 2}{\rm e}^{\sigma}\rho^{a b}\rho^{c d}
 (D_{+}\rho_{a c})(D_{-}\rho_{b d})  \nonumber\\
& & +{\rm e}^{\sigma} R_2       
    +2h{\rm e}^{\sigma}\Big\{
  D_{-}^{2}\sigma +{1\over 2} (D_{-}\sigma)^{2} 
 + {1\over 4}\rho^{a b}\rho^{c d}
   (D_{-}\rho_{a c})(D_{-}\rho_{b d}) \Big\}  
    \Big]                       \nonumber\\
& & + {\rm surface} \ {\rm terms}.   \label{react}     
\end{eqnarray}
Here $+$ and $-$ stands for $u$ and $v$, respectively, 
and $R_{2}$ is the scalar curvature of the fibre space $N_{2}$.
We summarize the notations below;
\begin{eqnarray}
& &F_{+-}^{\ \ a}=\partial_{+} A_{-} ^ { \ a}-\partial_{-}
  A_{+} ^ { \ a} - [A_{+}, A_{-}]_{\rm L}^{a}  \nonumber\\
& &\hspace{.95cm}=\partial_{+} A_{-} ^ { \ a}-\partial_{-}
  A_{+} ^ { \ a}-A_{+}^{\ c}\partial_{c}A_{-}^{\ a}
  +A_{-}^{\ c}\partial_{c}A_{+}^{\ a},   \label{field}\\
& &D_{\pm}\sigma = \partial_{\pm}\sigma
-[A_{\pm}, \sigma]_{\rm L}        \nonumber\\
& &\hspace{.95cm}=\partial_{\pm}\sigma
-A_{\pm}^{\ a}\partial_{a}\sigma
-\partial_{a}A_{\pm}^{\ a},         \label{get}\\
& &D_{\pm}\rho_{a b}=\partial_{\pm}\rho_{a b}
   - [A_{\pm}, \rho]_{{\rm L}a b} \nonumber\\
& &\hspace{1.25cm}=\partial_{\pm}\rho_{a b}
-A_{\pm} ^ { \ c}\partial_c \rho_{a b}
-(\partial_a A_{\pm} ^ { \ c})\rho_{c b}
-(\partial_b A_{\pm} ^ { \ c})\rho_{a c}
+(\partial_c A_{\pm} ^ { \ c})\rho_{a b},  \label{rhod}
\end{eqnarray}
where
$[A_{\pm}, \ast]_{\rm L}$ is the Lie derivative 
of $\ast$ along the vector fields 
$A_{\pm}:=A_{\pm}^{\ a}\partial_{a}$.
Each term in $I_{4}$ strongly suggests that
the integral should be interpreted as an action integral of 
a Yang-Mills type gauge theory defined on the 2-dimensional
base manifold $M_{1+1}$, interacting with the 2-dimensional  
field $\sigma$ and non-linear sigma field $\rho_{a b}$.
The associated local gauge symmetry is 
the built-in diff$N_{2}$ symmetry,
the group of the diffeomorphisms of the fibre space $N_{2}$.
It must be mentioned here that each term in (\ref{react}) is
manifestly diff$N_{2}$-invariant, and that the $y^{a}$-dependence 
is completely hidden in the Lie derivatives.
In this sense we may regard the fibre space $N_{2}$ as 
a kind of an {\it internal} space
as in Yang-Mills theories\cite{foot}.

Apart from the eight equations of motion that follow 
from (\ref{react}) by the variations, however, 
there are two additional equations we have 
to consider, which are associated with the gauge fixing of 
the 2-dimensional metric to the form (\ref{poly}). 
These equations that follow by varying the Einstein-Hilbert action 
{\it before} we impose the gauge fixing condition, 
turn out to be two of the four Einstein's constraints\cite{summary}.
They are found to be
\begin{eqnarray}
&(a)&\hspace{.5cm} {\rm e}^{\sigma} D_{+}D_{-}\sigma 
+ {\rm e}^{\sigma} D_{-}D_{+}\sigma  
+ 2{\rm e}^{\sigma} (D_{+}\sigma)(D_{-}\sigma)
- 2{\rm e}^{\sigma}(D_{-}h)(D_{-}\sigma)  \nonumber\\
& &\hspace{.5cm} - {1\over 2}{\rm e}^{ 2 \sigma}\rho_{a b}
   F_{+-}^{\ \ a}F_{+-}^{\ \ b}
- {\rm e}^{\sigma} R_{2}
- h {\rm e}^{\sigma} \Big\{
(D_{-}\sigma)^{2} 
-{1\over 2}\rho^{a b}\rho^{c d} 
 (D_{-}\rho_{a c})(D_{-}\rho_{b d})\Big\}=0, \label{aa}\\
&(b)&\hspace{.5cm}-{\rm e}^{\sigma} D_{+}^{2}\sigma 
- {1\over 2}{\rm e}^{\sigma}(D_{+}\sigma)^{2}
-{\rm e}^{\sigma}(D_{-}h) (D_{+}\sigma)
+{\rm e}^{\sigma}(D_{+}h)(D_{-}\sigma) \nonumber\\
& &\hspace{.5cm}+2h {\rm e}^{\sigma}(D_{-}h)(D_{-}\sigma) 
+{\rm e}^{\sigma}F_{+-}^{\ \ a}\partial_{a}h
-{1\over 4}{\rm e}^{\sigma}\rho^{a b}\rho^{c d} 
 (D_{+}\rho_{a c})(D_{+}\rho_{b d})
+\partial_{a}\Big( \rho^{a b}\partial_{b}h \Big) \nonumber\\
& &\hspace{.5cm}+h\Big\{ - {\rm e}^{\sigma} (D_{+}\sigma) 
  (D_{-}\sigma)
+{1\over 2}{\rm e}^{\sigma}\rho^{a b}\rho^{c d} (D_{+}\rho_{a c})
    (D_{-}\rho_{b d})
+{1\over 2}{\rm e}^{2\sigma}\rho_{a b}F_{+-}^{\ \ a}F_{+-}^{\ \ b}
+{\rm e}^{\sigma}R_{2} \Big\} \nonumber\\
& &\hspace{.5cm}+h^{2}{\rm e}^{\sigma}\Big\{
(D_{-}\sigma)^{2}
-{1\over 2}\rho^{a b}\rho^{c d}
(D_{-}\rho_{a c}) (D_{-}\rho_{b d})\Big\}=0.  \label{bb}
\end{eqnarray}
Together with the above equations, the ten Einstein's equations 
are given by
\begin{eqnarray}
&(c)&\hspace{.5cm}2{\rm e}^{\sigma}(D_{-}^{2}\sigma) +
{\rm e}^{\sigma} (D_{-}\sigma)^{2}  
    + {1\over 2}{\rm e}^{\sigma}\rho^{a b}\rho^{c d} (D_{-}\rho_{a c})
    (D_{-}\rho_{b d})=0,                 \label{cc}\\
&(d)&\hspace{.5cm}D_{-}\Big( {\rm e}^{2\sigma}
\rho_{a b}F_{+-}^{\ \ b}\Big)
- {\rm e}^{\sigma}\partial_{a}(D_{-}\sigma) 
- {1\over 2}{\rm e}^{\sigma}\rho^{b c}\rho^{d e}
    (D_{-}\rho_{b d})(\partial_{a}\rho_{c e}) 
+ \partial_{b} \Big(
{\rm e}^{ \sigma}\rho^{b c}D_{-}\rho_{a c} \Big)\nonumber\\
& &\hspace{.5cm}=0,                   \label{dd}\\
&(e)&\hspace{.5cm}-D_{+}\Big( {\rm e}^{2\sigma}
\rho_{a b}F_{+-}^{\ \ b}\Big) 
-{\rm e}^{\sigma}\partial_{a} (D_{+}\sigma )
   -{1\over 2}{\rm e}^{\sigma}\rho^{b c}\rho^{d e}
   (D_{+}\rho_{b d})(\partial_{a}\rho_{c e})  \nonumber\\
& &\hspace{.5cm}+\partial_{b}\Big(  
{\rm e}^{ \sigma}\rho^{b c}D_{+}\rho_{a c} \Big)
+2h{\rm e}^{\sigma}\partial_{a}(D_{-}\sigma)   
+h{\rm e}^{\sigma}\rho^{b c}\rho^{d e}
   (D_{-}\rho_{b d})(\partial_{a}\rho_{c e}) 
+2{\rm e}^{\sigma}\partial_{a}(D_{-}h)    \nonumber\\
& &\hspace{.5cm}-2\partial_{b}\Big(h  
{\rm e}^{ \sigma}\rho^{b c}D_{-}\rho_{a c} \Big)
=0,                              \label{ee}\\
&(f)&\hspace{.5cm}-2 {\rm e}^{ \sigma}D_{-}^{2} h 
-2{\rm e}^{ \sigma} (D_{-}h)(D_{-}\sigma) 
+ {\rm e}^{ \sigma}D_{+}D_{-}\sigma 
+ {\rm e}^{ \sigma}D_{-}D_{+}\sigma        
+ {\rm e}^{ \sigma} (D_{+}\sigma)(D_{-}\sigma) \nonumber\\  
& &\hspace{.5cm}+ {1\over 2}{\rm e}^{ \sigma}\rho^{a b}\rho^{c d} 
  (D_{+}\rho_{a c})(D_{-}\rho_{b d})   
+ {\rm e}^{2 \sigma}\rho_{a b}
    F_{+-}^{\ \ a}F_{+-}^{\ \ b}
 -2h{\rm e}^{ \sigma} \Big\{
   D_{-}^{2} \sigma +{1\over 2}(D_{-}\sigma)^{2} \nonumber\\
& &\hspace{.5cm}+{1\over 4}\rho^{a b}\rho^{c d}
   (D_{-}\rho_{a c})(D_{-}\rho_{b d})\Big\}=0,  \label{ff}\\
&(g)&\hspace{.5cm}h\Big\{ {\rm e}^{\sigma} D_{-}^{2} \rho_{ab} 
- {\rm e}^{\sigma}\rho^{c d}(D_{-}\rho_{a c})(D_{-}\rho_{b d}) 
+{\rm e}^{\sigma}(D_{-}\rho_{a b})(D_{-}\sigma) \Big\}  \nonumber\\
& &\hspace{.5cm}-{1\over 2}{\rm e}^{\sigma} \Big( 
D_{+}D_{-}\rho_{a b} + D_{-}D_{+}\rho_{a b} \Big) 
+{1\over 2}{\rm e}^{\sigma} \rho^{c d}\Big\{ 
(D_{-}\rho_{a c})(D_{+}\rho_{b d}) 
+(D_{-}\rho_{b c})(D_{+}\rho_{a d}) \Big\}  \nonumber\\
& &\hspace{.5cm}-{1\over 2}{\rm e}^{\sigma}\Big\{ 
(D_{-}\rho_{a b})(D_{+}\sigma)
+(D_{+}\rho_{a b})(D_{-}\sigma)  \Big\}  \nonumber\\
& &\hspace{.5cm}
 +{\rm e}^{\sigma}(D_{-}\rho_{a b})(D_{-}h)                
+{1\over 2}{\rm e}^{2 \sigma}\rho_{a c}\rho_{b d}
 F_{+-}^{\ \ c}F_{+-}^{\ \ d}     
-{1\over 4}{\rm e}^{2 \sigma}\rho_{a b}
 \rho_{c d}F_{+-}^{\ \ c}F_{+-}^{\ \ d}=0, \label{gg}
\end{eqnarray}
which are the equations of motion of $h$, $A_{+}^{\ a}$, 
$A_{-}^{\ a}$, $\sigma$, and $\rho_{a b}$, respectively. 

Now we shall obtain the spherically symmetric
vacuum solution of the Einstein's equations by solving the above
equations. For this purpose
we need to write down the spherically symmetric
line element in the form (\ref{pol})\cite{synge}. Let us recall that
the spherical symmetry with respect to a given observer means
that the metric is independent of the orientation 
at each point on the worldline $C$ 
of that observer (see Fig.1).
Let $\vartheta$ and $\varphi$ be the angular coordinates 
that define the orientation at that point. 
Then, due to the spherical symmetry, 
it suffices to consider the 2-dimensional subspace
defined by $\vartheta={\rm constant}$ and $\varphi={\rm constant}$.
Let $(u,v)$ be the coordinates of an arbitrary event 
$E$ in this subspace, which we introduce as follows;
(a) Given an event $E$, draw a past-directed null geodesic 
from $E$ cutting the worldline $C$ at $P$. The coordinate $v$ is
defined as the affine distance of the event $E$ from $P$ 
along the null geodesic.
(b) The coordinate $u$ measures the location of the event 
$P$ from a certain reference point $O$ along the worldline $C$.
The affine parameter $v$ has the coordinate freedom 
\begin{equation}
v \longrightarrow v'= A(u) v + B(u),     \label{aff}
\end{equation}
on each null hypersurface defined by $u={\rm constant}$.
Also there is a reparametrization invariance 
\begin{equation}
u \longrightarrow u'=f(u),
\end{equation}
where $f(u)$ is an arbitrary function of $u$.
Notice that the equation $du=0$ defines a null geodesic 
in the $(u,v)$-subspace. Choosing $A=1$ and $B=0$ in 
(\ref{aff}), we can write the metric of this subspace 
as a product of $du$ and $dv + h(u,v) du$, where 
$h$ is an arbitrary function of $(u,v)$.
Therefore the metric of 
the spherically symmetric spacetime is given by
\begin{equation}
ds^2= -2 du dv -2 h(u,v) du^{2} 
+H(u,v)(d\vartheta^2+\sin^2\vartheta d\varphi^2), \label{spher}
\end{equation}
which we recognize as the spherically symmetric line element 
written in the form (\ref{pol}), if we identify
\begin{math}
 y^{a}=(\vartheta, \varphi).
\end{math}
Here the fibre space $N_{2}$ is a two sphere $S_{2}$
of radius $H^{1/2} (H > 0)$, whose scalar curvature 
is given by
\begin{equation}
R_{2}=-{2\over H}.
\end{equation}
If we compare (\ref{spher}) with (\ref{pol}), we find that 
\begin{eqnarray}
& &A_{\pm}^{\ \vartheta}=A_{\pm}^{\ \varphi}=0, \nonumber\\
& &\rho_{\vartheta \vartheta}
={1 \over {\rm sin} \vartheta}, \hspace{.5cm} 
\rho_{\varphi \varphi}= {\rm sin} \vartheta, \hspace{.5cm} 
\rho_{\vartheta \varphi}= 0, \nonumber\\
& &{\rm e}^{\sigma}=H \ {\rm sin} \vartheta.
\end{eqnarray}
Notice that the diff$N_{2}$-covariant derivatives $D_{\pm}$
reduce to $\partial_{\pm}$
since $A_{\pm}^{\ a}$ become zero.
Then the Einstein's equations (\ref{aa}), 
$\cdots$, (\ref{gg}) become 
\begin{eqnarray}
&(a)&  \hspace{.5cm} \partial_{+}\partial_{-}\sigma 
 + \partial_{-}\partial_{+}\sigma  
 + 2(\partial_{+}\sigma) (\partial_{-}\sigma)
 -2 (\partial_{-}\sigma)(\partial_{-}h)
 +{2\over H} - h (\partial_{-}\sigma)^{2}=0, \label{a}\\
&(b)&  \hspace{.5cm} \partial^{2}_{+}{\sigma} 
 +{1\over 2} (\partial_{+}{\sigma})^{2}
 +(\partial_{+}{\sigma})(\partial_{-}h)
 -(\partial_{-}{\sigma})(\partial_{+}h)
 -2h(\partial_{-}{\sigma}) (\partial_{-}h)   \nonumber\\
& &\hspace{.5cm} +h \Big\{
 (\partial_{+}{\sigma})(\partial_{-}{\sigma})
 +{2\over H} \Big\}
 -h^{2}(\partial_{-}{\sigma})^{2}=0, \label{b}  \\
&(c)& \hspace{.5cm}2 \partial^{2}_{-}{\sigma} 
 + (\partial_{-}{\sigma})^{2}=0,  \label{c}\\
&(d)& \hspace{.5cm}\partial_{a}\partial_{-}
 \sigma=0,                          \label{d}\\
&(e)& \hspace{.5cm}\partial_{a}\partial_{+}\sigma  
 -2h \partial_{a}\partial_{-}\sigma 
 -2 \partial_{a}\partial_{-}h=0,     \label{e}\\
&(f)&  \hspace{.5cm}\partial_{+}\partial_{-}\sigma 
 +\partial_{-}\partial_{+}\sigma 
 +(\partial_{+}\sigma) (\partial_{-}\sigma)
 -2 \partial^{2}_{-}h 
 -2 (\partial_{-}h)(\partial_{-}\sigma)=0,\label{f}\\
&(g)&  \hspace{.5cm}0=0.              \label{g}
\end{eqnarray}
respectively. 
Let us integrate Eq. (\ref{c}) first. It can be written as
\begin{equation}
2\partial_{-}X + X^{2}=0,           \label{ex}
\end{equation}
where
$X:=\partial_{-}{\sigma}$. 
Solving this equation, we find that
\begin{equation}
X={2\over v + 2F},        \label{second}
\end{equation}
where $F$ is an arbitrary function
of $(u, \vartheta, \varphi)$. Therefore $\sigma$ becomes 
\begin{eqnarray}
& &\sigma = 2{\rm ln}\ (v + 2 F)
     + G              \nonumber\\ 
& & \hspace{.38cm} = {\rm ln} \ H + {\rm ln}\ {\rm sin}\vartheta,
\end{eqnarray}
where $G$ is another arbitrary function of $(u, \vartheta, \varphi)$.
Choosing $F=0$ and  $G={\rm ln}\ {\rm sin}\vartheta$, we find that
\begin{eqnarray}
& &\sigma=2 {\rm ln}\  v + {\rm ln}\ {\rm sin}\vartheta, \nonumber\\
& &H=v^{2},                              \label{easy}
\end{eqnarray}
from which it follows that
\begin{equation}
\partial_{-}{\sigma} = {2\over v}, \hspace{.5cm} 
\partial_{+}{\sigma}=0.               \label{simple}
\end{equation}
Then Eqs. (\ref{d}) and (\ref{e}) are trivially satisfied, 
and the remaining equations become
\begin{eqnarray}
&(a)& \hspace{.5cm} 2\partial_{-} h
  + {2\over v}\ h - {1\over v}=0,      \label{newa} \\
&(f)& \hspace{.5cm} \partial^{2}_{-} h 
  + {2\over v}\ (\partial_{-} h)=0.     \label{newf}
\end{eqnarray}
Notice that the Eq. (\ref{b}) becomes identical to 
Eq. (\ref{newa}), and that Eq. (\ref{newf})
is a ``trivial'' equation since it results by taking 
a derivative of Eq. (\ref{newa}) with respect to $v$. 
Therefore we need to solve Eq. (\ref{newa}) only.
Assuming the asymptotic flatness at the null infinity
$v\rightarrow \infty$,
we find that $h$ is given by
\begin{equation}
2h=1-{2m\over v},                     \label{eichi}
\end{equation}
where $m$ is a constant. 
Plugging (\ref{easy}) and (\ref{eichi}) into (\ref{spher}),
the spherically symmetric solution of the vacuum 
Einstein's equations is given by 
\begin{equation}
ds^2= -2dudv-( 1-{2m\over v}) du^2
 + v^{2} (d\vartheta^2+\sin^2\vartheta d\varphi^2). \label{sol}
\end{equation}
Thus we found the Schwarzschild solution 
using the Kaluza-Klein variables adapted
to the (2,2)-fibration of 4-dimensional spacetimes. 
Notice that the metric (\ref{sol}) is independent of
$u$ (as well as $\vartheta$ and $\varphi$), which tells us
that $u$ is the Killing time, as implied 
by the Birkhoff's theorem. 

There are a few possible applications of this formalism.
First, this formalism provides 
a natural 2-dimensional framework for a conventional gauge theory 
description of general relativity of 4-dimensional spacetimes,
where the local gauge symmetry is diff$N_{2}$, the infinite 
dimensional group of the diffeomorphisms of 
the 2-dimensional ``auxiliary'' space. 
This enables us to explore certain canonical aspects of 
the theory, such as constructing physical observables for instance,
using the gauge invariant quantities. Probably one could
also use relevant 2-dimensional field theoretic methods in studying
4-dimensional spacetime physics in this formalism\cite{w}.

Second, we expect this formalism to fit most naturally
the studies of gravitational waves, since
the physical degrees of freedom of gravitational waves reside
precisely in the non-linear sigma field $\rho_{ab}$\cite{sachs}.
It is also an interesting problem to examine exact solutions of 
the Einstein's equations in the light of this formalism, and
interpret them from the 2-dimensional gauge theory perspective.
For instance, the Schwarzschild spacetime in this Letter
corresponds
to the ``vacuum'' configuration, in the sense that the gauge
fields $A_{\pm}^{\ a}$ are identically zero. 

Third, this formalism should be compared with 
the lightcone cut formalism\cite{kozameh}
of C. Kozameh and E.T. Newman,
where the lightcone cuts are the {\it master} fields 
defined at the null infinity of asymptotically flat spacetimes.
In that formalism the lightcone cuts at the null infinity 
are constructed using the Bondi coordinates, in which 
the metric assumes the form (\ref{pol}), and $N_{2}$ becomes 
$S_{2}$. The difference is that our formalism
is valid for a generic spacetime, at least locally, 
whereas the lightcone cut formalism
depends heavily on the asymptotic structure at infinity.

Finally, the self-dual Einstein's equations have been studied 
extensively from diverse points of view, and each approach has
its own advantage. Surely this Kaluza-Klein formalism 
will add one more to the list.

This work is supported in part by Korea Science and Engineering 
Foundation (95-0702-04-01-3) and by the SRC program of 
SNU-CTP of KOSEF. It is also supported by non-directed research 
fund, Korea Research Foundation, 1995 (04-D-0073).

\bigskip\goodbreak

\pagebreak

\begin{figure}
\centerline{\epsfxsize=10cm\epsfbox{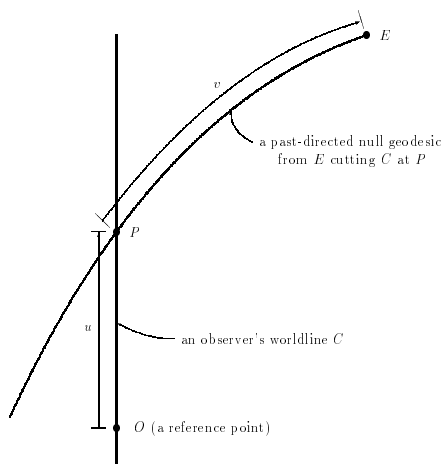}}
\vspace{-6cm}
\caption{The construction of the coordinates $(u,v)$ in the line 
element (\ref{spher}) assuming the spherical symmetry about an 
observer. Here the angular coordinates ($\vartheta$, $\varphi$) are
suppressed.}
\label{fig1}
\end{figure}


\begin{thebibliography}{99}
\bibitem{foot}{Integration over the $y^{a}$-coordinates 
might be viewed as an infinite dimensional
generalization of taking finite dimensional trace
over the fibre space indices
in Yang-Mills theories with some finite dimensional gauge
symmetries.}
\end{thebibliography}
\end{document}